\documentclass[journal]{IEEEtran}
\usepackage{url}
\usepackage[caption=false,font=footnotesize]{subfig}
\usepackage{dsfont}
\usepackage{todonotes}

\usepackage{amsmath,amssymb,amsthm}
\usepackage[printonlyused,nolist]{acronym}
\usepackage{xspace}
\usepackage{bbm}
\usepackage{booktabs}
\usepackage{graphicx}
\usepackage{tikz}
\usetikzlibrary{positioning,fit,intersections,calc}
\usepackage{pgfplots}
\usepackage{trsym}
\usepackage{siunitx}
\usepackage{bm}
\usepackage[normalem]{ulem}

\pgfplotsset{compat=newest}

\usepackage{amsmath,amssymb,amsthm}
\usepackage{xspace}
\usepackage{bbm}
\usepackage{booktabs}
\usepackage{graphicx}
\usepackage{tikz}
\usepackage{pgfplots}
\usepackage{trsym}
\usepackage{siunitx}

\newcommand{\setc}{\ensuremath{\mathcal{C}}\xspace}

\DeclareMathOperator{\entop}{\mathbb{H}}

\DeclareMathOperator{\miop}{\mathbb{I}}

\DeclareMathOperator*{\argmin}{argmin}

\newcommand{\vecc}{\boldsymbol{c}}

\newcommand{\vecp}{\boldsymbol{p}}

\newcommand{\vecs}{\boldsymbol{s}}

\newcommand{\vecu}{\boldsymbol{u}}

\newcommand{\vecv}{\boldsymbol{v}}

\newcommand{\vecx}{\boldsymbol{x}}

\newcommand{\mathh}{\boldsymbol{H}}

\newcommand{\matq}{\boldsymbol{Q}}

\newcommand{\matr}{\boldsymbol{R}}
\newcommand{\veczero}{\boldsymbol{0}}

\newcommand{\bpm}{\begin{pmatrix}}
\newcommand{\epm}{\end{pmatrix}}
\newcommand{\bbm}{\begin{bmatrix}}
\newcommand{\ebm}{\end{bmatrix}}

\definecolor{TUMBeamerYellow}    {rgb}{1.00,0.71,0.00}  
\definecolor{TUMBeamerOrange}    {rgb}{1.00,0.50,0.00}  
\definecolor{TUMBeamerRed}       {rgb}{0.90,0.20,0.09}  
\definecolor{TUMBeamerDarkRed}   {rgb}{0.79,0.13,0.25}  
\definecolor{TUMBeamerBlue}      {rgb}{0.00,0.60,1.00}  
\definecolor{TUMBeamerLightBlue} {rgb}{0.25,0.75,1.00}  
\definecolor{TUMBeamerGreen}     {rgb}{0.57,0.67,0.42}  
\definecolor{TUMBeamerLightGreen}{rgb}{0.71,0.79,0.51}  

\newcommand{\hw}{\textnormal{w}_{\textnormal{H}}}

\theoremstyle{plain}

\theoremstyle{remark}
\newtheorem{remark}{Remark}

\theoremstyle{definition}
\usepackage{mdframed}
\definecolor{examplegray}{rgb}{.95,.95,.95}
\newtheorem{mdexample}{Example}[section]

\title{Probabilistic Parity Shaping for Linear Codes}
\author{Georg B\"ocherer,~\IEEEmembership{Member,~IEEE}, Diego Lentner, Alessandro Cirino, Fabian Steiner,~\IEEEmembership{Student~Member,~IEEE}
\thanks{Georg B\"ocherer, Diego Lentner, and Alessandro Cirino are with the Mathematical and Algorithmic Sciences Lab, Huawei Technologies France S.A.S.U. Email: \texttt{georg.boecherer@ieee.org}, \texttt{diego.lentner@gmx.de}, \texttt{alessandro.cirino@studio.unibo.it}. Alessandro Cirino is also with the University of Bologna. Fabian Steiner is with the Institute for Communications Engineering, Technical University of Munich. Email: \texttt{fabian.steiner@tum.de}.
}}

\definecolor{TUMBeamerYellow}    {rgb}{1.00,0.71,0.00}  
\definecolor{TUMBeamerOrange}    {rgb}{1.00,0.50,0.00}  
\definecolor{TUMBeamerRed}       {rgb}{0.90,0.20,0.09}  
\definecolor{TUMBeamerDarkRed}   {rgb}{0.79,0.13,0.25}  
\definecolor{TUMBeamerBlue}      {rgb}{0.00,0.60,1.00}  
\definecolor{TUMBeamerLightBlue} {rgb}{0.25,0.75,1.00}  
\definecolor{TUMBeamerGreen}     {rgb}{0.57,0.67,0.42}  
\definecolor{TUMBeamerLightGreen}{rgb}{0.71,0.79,0.51}  

\definecolor{TUMBlau}      {rgb}{0.00,0.60,1.00}  

\usepackage{array}

\newcommand{\revgb}[1]{#1}

    \begin{acronym}
    \acro{ASI}{asymmetric information}
    \acro{AWGN}{additive white Gaussian noise}
    \acro{ADM}{amplitude distribution matcher}
    \acro{BPSK}{binary phase shift keying}
    \acro{DPC}{dirty paper coding}
    \acro{FEC}{forward error correction}
    \acro{LUT}{lookup table}
    \acro{PAS}{probabilistic amplitude shaping}
    \acro{PSEnc}{probabilistic shaping encoder}
    \acro{SDM}{syndrome distribution matcher}
    \acro{SMD}{symbol metric decoding}    
    \acro{BMD}{bit metric decoding}
    \acro{LDPC}{low-density parity-check}
    \acro{LLR}{log-likelihood ratio}
    \acro{LLPS}{linear layered probabilistic shaping}
    \acro{BER}{bit error rate}
    \acro{CCDM}{constant composition distribution matching}
    \acro{DM}{distribution matching}
    \acro{QAM}{quadrature amplitude modulation}
    \acro{ASK}{amplitude shift keying}
    \acro{OOK}{on-off-keying}
    \acro{OH}{overhead}
    \acro{SNR}{signal-to-noise-ratio}
    \acro{FER}{frame error rate}
    \acro{BICM}{bit-interleaved coded modulation}
    \acro{WLLN}{weak law of large numbers}
    \acro{DSP}{digital signal processing}
    \acro{BRGC}{binary reflected Gray code}
    \acro{PDM}{polarization division multiplexing}
    \acro{PPS}{probabilistic parity shaping}
    \acro{SD-FEC}{soft decision forward error correction}
    \acro{HD-FEC}{hard decision forward error correction}
    \acro{SSMF}{standard single-mode fiber}
    \acro{SMF}{single-mode fiber}
    \acro{IM}{intensity modulation}
    \acro{PMF}{probability mass function}
    \acro{PS}{probabilistic shaping}
    \acro{RV}{random variable}
    \acro{PDF}{probability density function}
    \acro{NB}{non-binary}
    \acro{GMI}{generalized mutual information}
    \acro{NGMI}{normalized generalized mutual information}
    \acro{MLC-MSD}{multilevel coding with multistage decoding}
    \acro{HD}{hard decision}
    \acro{BSC}{binary symmetric channel}
    \acro{CC}{constant composition}
    \acro{MB}{Maxwell-Boltzmann}
    \acro{SPC}{single parity check}
    \acro{SE}{spectral efficiency}
    \end{acronym}

\newcommand{\kinfo}{k_\textnormal{info}}
\newcommand{\rfec}{R_\textnormal{fec}}

\begin{document}

\maketitle

\begin{abstract}
Linear layered probabilistic shaping (LLPS) is proposed, an architecture for linear codes to efficiently encode to shaped code words. In the previously proposed probabilistic amplitude shaping (PAS) architecture, a distribution matcher (DM) maps information bits to shaped bits, which are then systematically encoded by appending uniformly distributed parity bits. LLPS extends PAS by probabilistic parity shaping (PPS), which uses a syndrome DM to calculate shaped parity bits. LLPS enables the transmission with any desired distribution using linear codes, furthermore, by LLPS, a given linear code with rate $\rfec$ can be operated at any rate $R\leq\rfec$ by changing the distribution. LLPS is used with an LDPC code for dirty paper coding against an interfering BPSK signal, improving the energy efficiency by 0.8~dB.
\end{abstract}

\section{Introduction}
\label{sec:intro}

Communication channels often have non-uniform capacity-
achieving input distributions, which has been the main mo-
tivation for \ac{PS}, i.e., the development
of practical transmission schemes that use non-uniform input
distributions. Many different \ac{PS} schemes have been proposed
in literature, see, e.g., the literature review in \cite[Sec.~II]{bocherer2015bandwidth}. Probabilistic amplitude shaping (PAS)\acused{PAS}~\cite{bocherer2015bandwidth} uses \ac{DM} to map information bits to shaped bits, which are then systematically encoded to append uniformly distributed parity bits. \ac{PAS} integrates with any linear \ac{FEC} code. For higher-order modulation for the \ac{AWGN} channel, \ac{PAS} is capacity-achieving~\cite[Sec.~10.3]{bocherer2018principles},\cite{amjad2018information} and has found wide applications for optical~\cite{bocherer2019probabilistic}, wired~\cite{iannone2018increasing}, and wireless~\cite{gultekin2017constellation} transmission.

However, there are important cases where optimal transmission requires shaped parities~\cite[Remark~3]{bocherer2019probabilistic}, examples include intensity modulation~\cite{eriksson2017probabilistically} and \ac{OOK}. A time-sharing based shaping scheme (\emph{sparse-dense-transmission}) for \ac{OOK} was presented in \cite{git_protograph-based_2019}, while an implementation for polar codes is shown in \cite{wiegart2019_ook_polar}.

The layered \ac{PS} \revgb{random code ensemble} introduced in \cite{bocherer2017achievable,bocherer2018principles,bocherer2019probabilistic} suggests that encoding to shaped parities is indeed possible, in particular, it suggests that for linear codes of length $n$, dimension $k$, and rate $\rfec=k/n$, we can encode to code words with distribution $P_B$ at rate
\begin{align}
R=\left[\entop(B)-(1-\rfec)\right]^+
\end{align}
where $\entop(B)$ denotes the entropy of $B$. However, no efficient encoding algorithm is known, see e.g., \cite{eriksson2017probabilistically},\cite[Remark~3]{bocherer2019probabilistic}, which means that encoding has to be done by a \ac{LUT} with $2^{Rn}$ entries~\cite[Sec.~II-E]{bocherer2019probabilistic}, which is prohibitively large already for short codes.

\emph{Contribution:} In this work, we suggest \ac{LLPS}, which extends \ac{PAS} by \ac{PPS}, which can be realized by a \ac{SDM}. For any binary linear code of length $n$ and dimension $k$, the \ac{SDM} can be realized by calculating online a set of size $2^\ell$, where
\begin{align}
\ell\approx (n-k)\left(\frac{1}{\entop(B)}-1\right)
\end{align}
or by calculating offline a \ac{LUT} of size $2^{n-k}$. The numbers $\ell$ and $n-k$ can be much smaller than \revgb{$Rn$}. 

We apply \ac{LLPS} to coding against an interfering \ac{BPSK} signal that is known in advance to the transmitter but not to the receiver. This is an instance of the class of channels considered by Gelfand and Pinsker in \cite{gelfand1980coding}, for which transmission schemes are often called \ac{DPC}, following~\cite{costa1982writing}. For an $n\approx 1000$ rate 1/2 \ac{LDPC} code, \ac{DPC} by \ac{LLPS} improves the energy efficiency by $\SI{0.8}{dB}$, with $\ell= 16$. Compared to a naive layered \ac{PS}, the size of the required \ac{LUT} is reduced from $2^{nR}\approx 2^{500}$ to $2^{16}$, which is significantly smaller.

\emph{Outline:} In Sec.~\ref{sec:preliminaries}, we briefly review systematic encoding, layered \ac{PS}, and \ac{PAS}. We introduce \ac{LLPS} in Sec.~\ref{sec:pps}. We then apply \ac{LLPS} to \ac{DPC} in Sec.~\ref{sec:dpc} and present numerical results. We conclude in Sec.~\ref{sec:conclusions}, pointing out future research directions.

\emph{Notation:} We denote random variables by capital letters, e.g., $X,Y$. We denote by $\entop(X)$ and $\entop(X|Y)$ the entropy of $X$ and $X$ conditioned on $Y$, respectively. $\miop(X;Y)$ denotes the mutual information of $X$ and $Y$.

\begin{figure}
\includegraphics[width=\columnwidth]{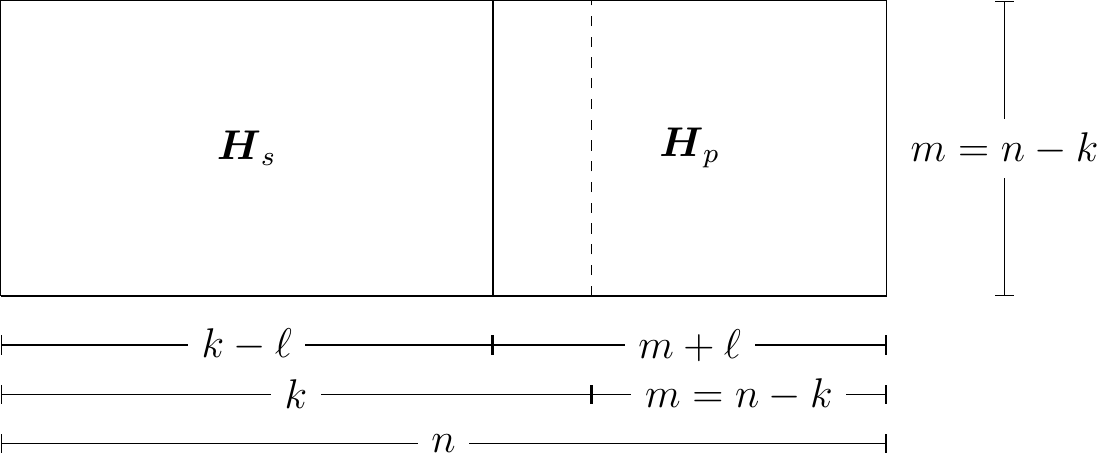}
\caption{\ac{PPS} divides the parity check matrix $\mathh$ into a syndrome former $\mathh_s$ and a non-square parity former $\mathh_p$.}
\label{fig:new H}
\includegraphics[width=\columnwidth]{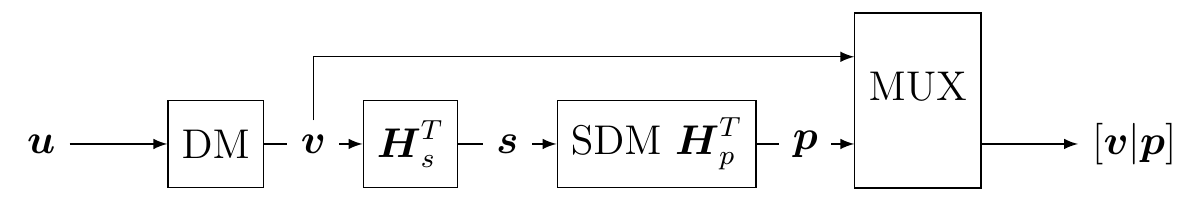}

\caption{\ac{LLPS} encoding is achieved in two steps: (1) syndrome $\vecs=\vecv\mathh_s^T$ is calculated from the (shaped) systematic part $\vecv$. (2) the \ac{SDM} calculates as parity part a shaped vector $\vecp$ with $\vecp\mathh_p^T=\vecs$. Without \ac{PPS}, i.e., when $\mathh_p$ is square and $\vecp=\vecs(\mathh_p^T)^{-1}$, the \ac{PAS}~\cite{bocherer2015bandwidth} architecture is recovered.
}
\label{fig:llps}
\end{figure}
\section{Preliminaries}
\label{sec:preliminaries}
\begin{figure}
\includegraphics[width=\columnwidth]{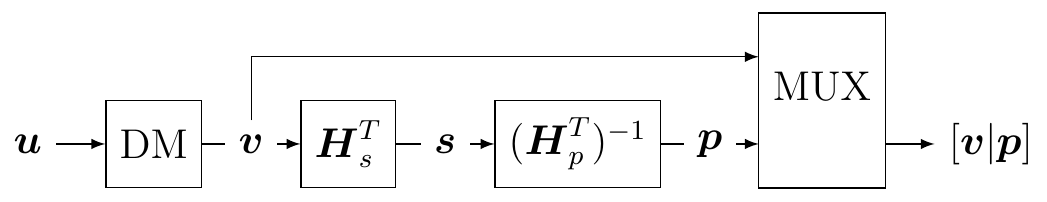}
\caption{\ac{PAS} \cite{bocherer2015bandwidth}.}
\label{fig:pas}
\end{figure}

\subsection{Systematic Encoding}
\label{subsec:systematic encoding}
Consider an $(n,k)$ binary linear code $\setc$ with block length $n$ and dimension $k$ and define $m=n-k$. We represent the code by an $m\times n$ parity check matrix $\mathh$, i.e.,
\begin{align}
\setc=\left\{\vecc\in\{0,1\}^n\colon \vecc\mathh^T=\veczero\right\}.\label{eq:code by H}
\end{align}
The code rate is $\rfec = \frac{k}{n}$. Suppose that $\mathh$ decomposes as
\begin{align}
\mathh=[\mathh_s|\mathh_p]
\end{align}
where $\mathh_s$ is $m\times k$ and $\mathh_p$ is $m\times m$ and has full rank. Then a length $k$ vector $\vecv$ can be systematically encoded into the codeword $[\vecv|\vecp]$ in two steps
\begin{enumerate}
\item Calculate the syndrome $\vecs=\vecv\mathh_s^T$.
\item Calculate the parity bits $\vecp=\vecs(\mathh_p^T)^{-1}$.
\end{enumerate}
Note that
\begin{align}
[\vecv|\vecp]\mathh^T=\vecv\mathh_s^T+\vecp\mathh_p^T=\vecs+\vecs=\veczero
\end{align}
that is, by \eqref{eq:code by H}, $[\vecv|\vecp]$ is indeed a codeword.

\subsection{Linear Codes and Shaping}

Consider a memoryless binary input channel $p_{Y|B}$. By \cite{bocherer2017achievable}, correct decoding is possible if the overhead $1-\rfec$ fulfills
\begin{align}
1-\rfec>\entop(B|Y).
\end{align}
To relate code parameters to information measures, we consider (hypothetical) \emph{ideal} codes with $1-\rfec=\entop(B|Y)$. The number of check equations of an ideal linear code is then given by
\begin{align}
m=n-k&=n(1-\rfec)\label{eq:m=1-R}\\
&=n\entop(B|Y).
\end{align}
\subsection{\ac{PAS}}
In \ac{PAS} (see Fig.~\ref{fig:pas}), length $\kinfo\leq k$ information bits $\vecu$ are mapped by a \ac{DM} to $k$ shaped bits $\vecv$ following the distribution $P_V$. The shaped bits $\vecv$ are then systematically encoded to the codeword $[\vecv|\vecp]$, as described in Sec.~\ref{subsec:systematic encoding}. Consequently, the transmitted codeword has $k$ shaped bits $\vecv$ and $m$ unshaped parity bits $\vecp$ with the uniform distribution $P_U$. In higher-order modulation, the partially shaped codeword can be used for optimal signaling by using the shaped bits to address amplitudes and the unshaped bits to address signs~\cite{bocherer2015bandwidth}.

The number of check equations $m$ of an ideal code is equal to the average uncertainty of \ac{PAS}, i.e.,
\begin{align}
m = n\left[\rfec\entop(V|Y)+(1-\rfec)\entop(U|Y)\right].\label{eq:pas m}
\end{align}
By \cite{schulte2016constant}, the ideal \ac{DM} has rate $\entop(V)$, so that
\begin{align}
R=\frac{\kinfo}{n}=\frac{\entop(V)k}{n}=\entop(V)\rfec.\label{eq:pas rate}
\end{align}
Combining \eqref{eq:pas m} and \eqref{eq:pas rate}, we get after some manipulations
\begin{align}
R&=\rfec\miop(V;Y)+(1-\rfec)\miop(U;Y).\label{eq:ideal pas rate 3}
\end{align}
We see that the time sharing realized by \ac{PAS} between shaped and unshaped channel inputs results in a time sharing achievable rate, which is in general suboptimal, by the concavity of mutual information in input distributions~\cite[Theorem~2.7.4]{cover2006elements}.

\section{Probabilistic Parity Shaping}
\label{sec:pps}
\revgb{We now develop \ac{PPS}, extending \ac{PAS} by shaped parity bits}. 
\subsection{Modified Systematic Encoding}
Consider Fig.~\ref{fig:new H} and Fig.~\ref{fig:llps}. As in Sec.~\ref{subsec:systematic encoding}, we consider an $(n,k)$ binary linear code with a $m\times n$ check matrix. We again partition the check matrix into $\mathh=[\mathh_s|\mathh_p]$, however, we modify the size of $\mathh_s$ and $\mathh_p$ to $m\times(k-\ell)$ and $m\times(m+\ell)$, respectively. The systematic encoding is as follows:
\begin{enumerate}
\item For length $k-\ell$ vector $\vecv$, calculate the syndrome $\vecs=\vecv\mathh_s^T$.
\item Calculate $m+\ell$ parity bits $\vecp$ by solving
\begin{align}
\vecp\colon \vecp\mathh_p^T=\vecs.
\end{align}
\end{enumerate}
Since $\mathh_p$ is $m\times(m+\ell)$, the condition $\vecp\mathh_p^T=\vecs$ is fulfilled by many different solutions $\vecp$, consequently, we can choose the parity bits $\vecp$ subject to a shaping constraint. This is realized by an \ac{SDM}, which we discuss in more detail next.

\subsection{\ac{SDM}}
 
 For some cost function $f$, \revgb{e.g., the Hamming weight 
\begin{align}
\hw(\vecp)=\sum_i \bm{1}(p_i\neq 0)
\end{align}
}
where $\bm{1}(\text{true})=1$, $\bm{1}(\text{false})=0$, an \ac{SDM} takes as input the length $m$ syndrome $\vecs$ and outputs a solution of
\begin{align}
\vecp=&\argmin_{\vecp'\in\{0,1\}^{m+\ell}} f(\vecp')\label{eq:sdm objective}\\
\text{subject to}\quad&\vecp'\mathh_p^T=\vecs.\label{eq:sdm constraint}
\end{align}
We next detail an \ac{SDM} realization that calculates all feasible vectors by \eqref{eq:sdm constraint} and then outputs the best according to \eqref{eq:sdm objective}.
\revgb{Let 
\begin{align}
\setc_p=\left\{\vecx\in\{0,1\}^{\ell+m}\colon \vecx\mathh_p^T=\veczero\right\}
\end{align}
be the $\ell$-dimensional code defined by $\mathh_p$. The feasible vectors of \eqref{eq:sdm constraint} form the coset of $\setc_p$ given by
\begin{align}
\left\{\vecx+\tilde{\vecp}(\vecs)\left|\vecx\in\setc_p\right.\right\}
\end{align}
where $\tilde{\vecp}(\vecs)$ is some particular solution of \eqref{eq:sdm constraint} used as representative of the coset. For $\mathh_p=[\matq|\matr]$ with $\matr$ square and full rank, a convenient representative is given by
\begin{align}
\tilde{\vecp}(\vecs)=[\veczero|\vecs(\matr^T)^{-1}].
\end{align}
The set of feasible vectors can now be calculated efficiently as follows.
\begin{enumerate}
\item Calculate offline $\setc_p$ and store it in memory.
\item Calculate online $\tilde{\vecp}(\vecs)=[\veczero|\vecs(\matr^T)^{-1}]$. \item The set of solutions is $\{\setc_p+\tilde{\vecp}(\vecs)\}$.
\end{enumerate}
}

\subsection{Rate Matching by \ac{LLPS}}

Suppose we use \ac{LLPS} to encode into code words with distribution $P_B$. We realize the \ac{SDM} by using as cost function the cross entropy
\begin{align}
f(\vecp)=\frac{1}{m+\ell}\sum_{i=1}^{m+\ell}\log_2\frac{1}{P_B(p_i)}.
\end{align}
The ideal \ac{FEC} code has $1-\rfec=\entop(V|Y)$, the ideal \ac{DM} has rate $\kinfo/(k-\ell)=\entop(B)$, and the ideal \ac{SDM} has rate $m/(m+\ell)=\entop(B)$, which translates into the following equations
\begin{align}
m&=n\entop(B|Y)\\
\kinfo&=\entop(B)(k-\ell)\\
m&=\entop(B)(m+\ell).
\end{align}
We now have
\begin{align}
R=\frac{\kinfo}{n}&=\entop(B)\frac{k-\ell}{n}\\
&=\entop(B)\frac{k+m-m-\ell}{n}\\
&=\entop(B)-\frac{\entop(B)(m+\ell)}{n}\\
&=\entop(B)-\frac{m}{n}\\
&=\entop(B)-\entop(B|Y)=\miop(B;Y).
\end{align}
We conclude that \ac{LLPS} can operate at any rate between $0$ (for $\entop(V)=0$) and $\rfec$ (for $\entop(V)=1$), and with ideal components, \ac{LLPS} achieves the optimal achievable rate $\miop(B;Y)$.
\subsection{\ac{LLPS} Decoding}

The \ac{FEC} decoder calculates its decision $[\hat{\vecv}|\hat{\vecp}]$ from the information it is provided by demapper. Since the transmitted $[\vecv|\vecp]$ is a code word, no change of the decoder is required. The decoder throws away the parity bits $\hat\vecp$ and outputs the decision $\hat{\vecv}$. For this, the only information required by the decoder is the value of $\ell$.

\section{Dirty Paper Coding}
\label{sec:dpc}
\begin{figure}
\includegraphics[width=\columnwidth]{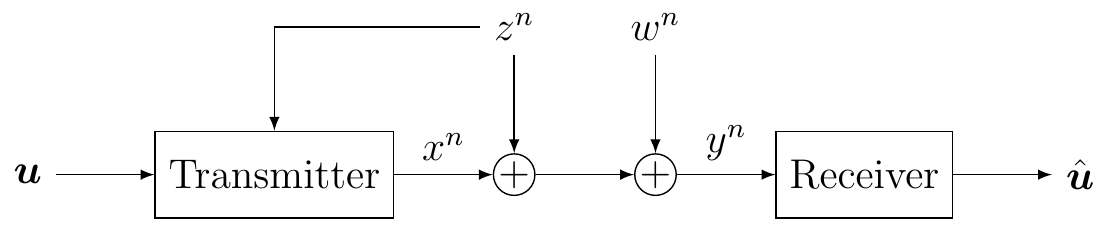}
\caption{Dirty paper coding scenario.}
\label{fig:dpc}
\end{figure}
We now apply \ac{LLPS} to a dirty paper coding scenario, where \acp{SDM} with small $\ell$, i.e., small computational cost, are sufficient to significantly improve the energy efficiency.

\subsection{Channel Setup}
We consider the scenario in Fig.~\ref{fig:dpc}. A binary sequence $b^n$ (not shown in Fig.~\ref{fig:dpc}) is mapped to a \ac{BPSK} signal $x^n$, which is transmitted. The received signal $y^n$ is the sum of the transmitted signal $x^n$, an interfering \ac{BPSK} signal $z^n$, and Gaussian noise $w^n$. The interfering signal $z^n$ is non-causally known to the transmitter, i.e., the binary sequence $b^n$ mapped to the transmitted signal $x^n$ is a function of the message $\vecu$ and the interfering signal $z^n$.  At time instance $i$, we have
\begin{align}
Y_i=\alpha x_{b_i}+\beta Z_i + W_i\label{eq:dpc_model}
\end{align}
where $W_i$, $i=1,\dotsc,n$ are independent and zero mean Gaussian with variance $\sigma^2$, where $z$ take values in $\{-1,1\}$, and where $x_0=-1$ and $x_1=+1$. The interference is  uniformly distributed, i.e., $P_Z(-1)=P_Z(1)=\frac{1}{2}$. We define the \ac{SNR} by $10\log_{10}(\alpha^2/\sigma^2)$~dB and we specify the strength of the interfering signal by $10\log_{10}(\beta^2/\alpha^2)$~dB.
\subsection{Reference Strategy: \revgb{Interference as Noise}}\label{sec:ref_int_as_noise}
The transmitter ignores the presence of $z^n$ and the receiver treats the interfering signal as noise. The achievable rate for this reference strategy is
\begin{align}
R&=\miop(B;Y)\nonumber\\
&\qquad\text{$B$ and $Z$ independent, $B$ uniformly distributed}.
\end{align}
The demapper calculates the \acp{LLR}
\begin{align}
L(y)=\log\frac{p_{Y|B}(y|0)}{p_{Y|B}(y|1)}
\end{align}
where
\begin{align}
p_{Y|B}(y|b)=\frac{1}{2}\left[p_W(y-\alpha x_b+\beta)+p_W(y-\alpha x_b-\beta)\right].
\end{align}
\subsection{\ac{DPC}}
\newcommand{\rdpc}{R_\text{dpc}}
Using \ac{DPC}  (see, e.g., \cite[Ch.~6]{kramer2007topics}) we can achieve the rate
\begin{align}
\rdpc=\miop(B;Y)-\miop(B;Z),\quad\text{$BZ\sim P_ZP_{B|Z}$}\label{eq:air dpc}
\end{align}
by transmitting $B$ according to $P_{B|Z}$. The demapper calculates the \ac{LLR}
\begin{align}
&p_{Y|B}(y|b)P_B(b)=\sum_{z\in\{-1,1\}}p_{Y|BZ}(y|bz)P_{BZ}(bz)\\
&=\sum_{z\in\{-1,1\}}p_{Y|BZ}(y|bz)P_{B|Z}(b|z)\frac{1}{2}\\
&=\sum_{z\in\{-1,1\}}p_W(y-\alpha x_b-\beta z)P_{B|Z}(b|z)\frac{1}{2}\label{eq:dpc_llr}.
\end{align}
\begin{remark}
We are considering the fixed bit-mapper $B\mapsto x_B$ with $x_0=-1$ and $x_1=+1$. By \cite[Ch.~6]{kramer2007topics},~\cite{lentner2018dirty}, in some cases, the \ac{DPC} achievable rate \eqref{eq:air dpc} can be further improved by using a time variant bit-mapper that depends on $z_i$.
\end{remark}

\subsection{\ac{LLPS} \ac{DPC} Encoder}

\begin{figure}
\centering
\footnotesize
\includegraphics{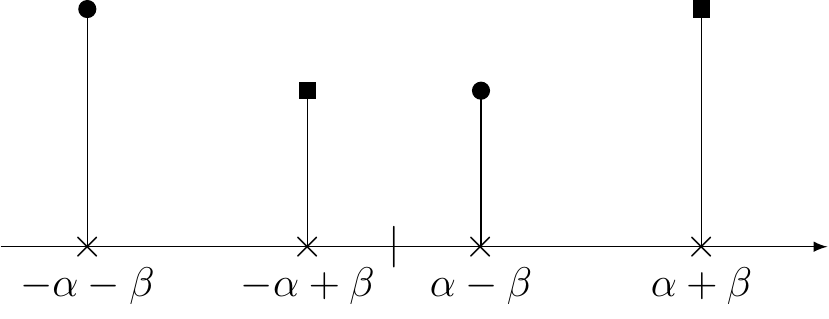}
\caption{The constellation of $\alpha X+\beta Z$ and the distribution $P_{X+Z}$ resulting from $P_{B|Z}$. The optimized distribution prefers the outer constellation points, since they can be detected more reliably. The circles indicate \text{$P_{X|Z}(\cdot|\num{-1})P_Z(\num{-1})$} and the squares indicate $P_{X|Z}(\cdot|1)P_Z(1)$.}
\label{fig:cstll}
\includegraphics[width=\columnwidth]{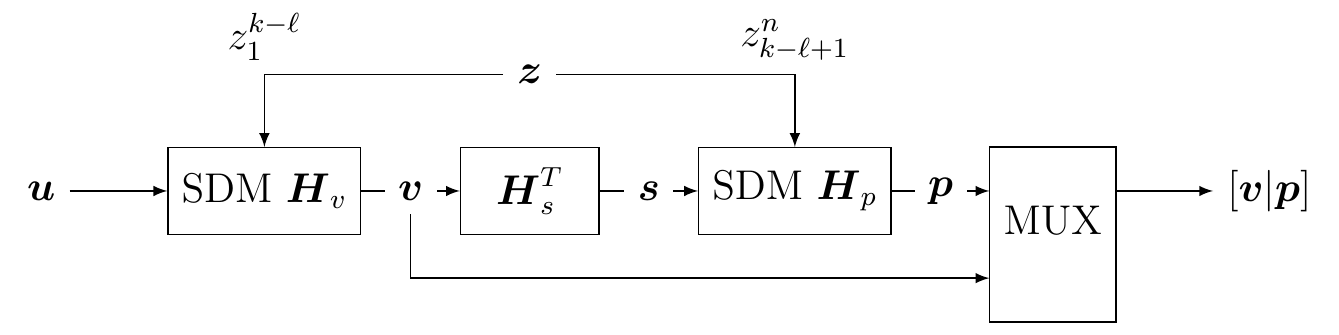}
\caption{\ac{LLPS} for dirty paper coding.}
\label{fig:dpc encoding}
\end{figure}
In Fig.~\ref{fig:dpc encoding}, we display the \ac{LLPS} for dirty paper coding. The \ac{DM} for the systematic part $\vecv$ is instantiated by a \ac{SDM} with matrix $\mathh_{\vecv}$, with an identity matrix to the right and entries at the left picked uniformly at random. Both \ac{SDM}s get provided the corresponding part of the interfering signal. In Fig.~\ref{fig:cstll}, we display the distributions that we obtained from optimizing \eqref{eq:air dpc}, see Sec.~\ref{subsec:numerical results}. {The figure suggests that the \ac{SDM}s should attempt to map 0 and 1 to the outermost signal points. Formally,} define the label of the interfering signal by
\begin{align}
a(z)=\begin{cases}
0&z=-1\\
1&z=1.
\end{cases}
\end{align}
Then, the \ac{SDM}s choose among the feasible vectors
\begin{align}
\vecv^*&=\argmin_{\vecv} \revgb{\hw(\vecv\oplus a_1^{k-\ell})}\\
\vecp^*&=\argmin_{\vecp} \revgb{\hw(\vecp\oplus a_{k-\ell+1}^n)}.
\end{align}


\subsection{Ideal \ac{LLPS} Rate}

By \eqref{eq:m=1-R}, we know that the ideal \ac{FEC} code has redundancy
\begin{align}
m=n-k=\entop(B|Y)n.
\end{align}
Furthermore, we know that an ideal \ac{SDM} has rate $\entop(B|Z)$. Thus, the rate of the \ac{DPC} transmitter is
\begin{align}
\frac{\entop(B|Z)(k-\ell)}{n}
&=\entop(B|Z)-\frac{\entop(B|Z)(n-k+\ell)}{n}\\
&=\entop(B|Z)-\frac{n-k}{n}\\
&=\entop(B|Z)-\entop(B|Y)\\
&=\miop(B;Y)-\miop(B;Z)
\end{align}
which recovers the achievable rate \eqref{eq:air dpc}.
\subsection{Numerical Results}
\label{subsec:numerical results}
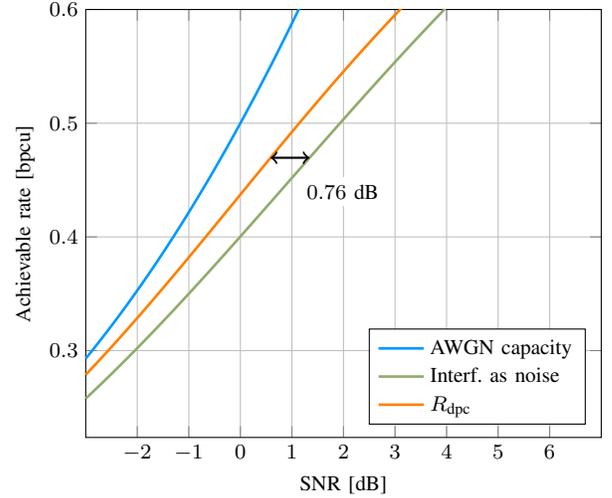
\begin{figure}
    \centering
    \footnotesize
    \begin{tikzpicture}
\begin{axis}[
xlabel={SNR [dB]},
ylabel={Achievable rate [bpcu]},
grid=both,
legend pos=south east,
legend cell align={left},
ymax=0.6,
xmin=-3,
xmax=7,
xtick={-2,-1,...,6}
]

\addplot[line width=1,TUMBeamerBlue] table[x=snr,y=r_cap] {data/rates.txt};
\addlegendentry{AWGN capacity};
\addplot[name path global=ref,line width=1,TUMBeamerGreen] table[x=snr,y=r_biawgn_int] {data/rates.txt};
\addlegendentry{Interf. as noise};
\addplot[name path global=dpc,line width=1,TUMBeamerOrange] table[x=snr,y=r_dpc] {data/rates.txt};
\addlegendentry{$R_{\text{dpc}}$};

\path[name path global=line] (-2,0.4696) -- (7,0.4696);
\path[name intersections={of=line and ref, name=p1}, name intersections={of=line and dpc, name=p2}];

\draw[<->,thick] let \p1=(p2-1), \p2=(p1-1) in (p1-1) -- (p2-1) node [below,midway,fill=white,yshift=-0.25cm,xshift=.7cm] {%
        \pgfplotsconvertunittocoordinate{x}{\x1}%
        \pgfplotscoordmath{x}{datascaletrafo inverse to fixed}{\pgfmathresult}%
        \edef\valueA{\pgfmathresult}%
        \pgfplotsconvertunittocoordinate{x}{\x2}%
        \pgfplotscoordmath{x}{datascaletrafo inverse to fixed}{\pgfmathresult}%
        \pgfmathparse{\pgfmathresult - \valueA}%
        \pgfmathprintnumber{\pgfmathresult} dB
};


\end{axis}
\end{tikzpicture}
    \caption{Information rates for different signaling schemes.}
    \label{fig:rates}
\end{figure}

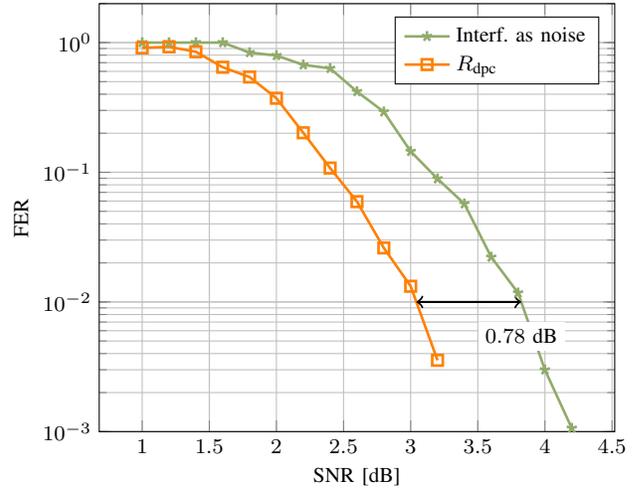
\begin{figure}
    \centering
    \footnotesize
    \begin{tikzpicture}
\begin{axis}[
xlabel={SNR [dB]},
ylabel={FER},
grid=both,
legend pos=north east,
legend cell align={left},
ymode=log,
xtick={1,1.5,...,4.5},
ymin=1e-3,
]

\addplot[name path global=ref,line width=1,TUMBeamerGreen,mark=star] table[x=snr,y=fer] {data/coded_results_ref1.txt};
\addlegendentry{Interf. as noise};
\addplot[name path global=dpc,line width=1,TUMBeamerOrange,mark=square,mark options={solid}] table[x=snr,y=fer] {data/coded_results_dpc1.txt};
\addlegendentry{$R_{\text{dpc}}$};

\path[name path global=line] (-2,1e-2) -- (7,1e-2);
\path[name intersections={of=line and ref, name=p1}, name intersections={of=line and dpc, name=p2}];

\draw[<->,thick] let \p1=(p2-1), \p2=(p1-1) in (p1-1) -- (p2-1) node [below,midway,fill=white,yshift=-0.25cm,xshift=.7cm] {%
        \pgfplotsconvertunittocoordinate{x}{\x1}%
        \pgfplotscoordmath{x}{datascaletrafo inverse to fixed}{\pgfmathresult}%
        \edef\valueA{\pgfmathresult}%
        \pgfplotsconvertunittocoordinate{x}{\x2}%
        \pgfplotscoordmath{x}{datascaletrafo inverse to fixed}{\pgfmathresult}%
        \pgfmathparse{\pgfmathresult - \valueA}%
        \pgfmathprintnumber{\pgfmathresult} dB
};

\end{axis}
\end{tikzpicture}
    \caption{Finite length simulation results.}
    \label{fig:coded_results1}
\end{figure}

In Fig.~\ref{fig:rates}, we show achievable rates for the considered \ac{DPC} setup. The blue curve provides the reference for the interference-free scenario assuming Gaussian signaling. The orange and green curve represent the case with interference and $10\log_{10}(\beta^2/\alpha^2)=\SI{-5}{dB}$. We observe that the orange \ac{LLPS} \ac{DPC} curve gains \SI{0.76}{dB} over the reference scheme, which treats interference as noise (see Sec.~\ref{sec:ref_int_as_noise}). The  employed non-uniform distribution $P_{B|Z}$ is obtained by maximizing $\rdpc$ in~\eqref{eq:air dpc}. 
In Fig.~\ref{fig:coded_results1}, we show finite length simulation results that target a transmission rate of $R = \SI{0.4697}{bits/channel~ use}$ (bpcu). The interference-as-noise scheme uses a rate $1/2$ Wimax code~\cite{ieee_wimax} with blocklength $n = \SI{1152}{bits}$, which is shortened by 66 bits to obtain the desired spectral efficiency. The \ac{LLPS} \ac{DPC} scheme uses a rate 1/2 Wimax code with blocklength $n = \SI{1056}{bits}$. The distribution employed for calculating the decoder soft information in \eqref{eq:dpc_llr} is 
\[
P_{B|Z} = \bbm 
    P_{B|Z}(0|\num{-1}) &   P_{B|Z}(0|1)\\
    P_{B|Z}(1|\num{-1})  &  P_{B|Z}(1|1)\\
\ebm = \bbm 
    0.6037 &   0.3963\\
    0.3963  &  0.6037\\
\ebm.
\]
The outer \ac{SDM} has rate  $\kinfo/(k-\ell) = 496/(528-16)=\SI{0.9688}{bits}$, and the inner \ac{SDM} has rate $m/(m+\ell) = \SI{0.9706}{bits}$. One hundred belief propagation iterations are performed. We observe gains of about \SI{0.8}{dB} in Fig.~\ref{fig:coded_results1}, recovering the asymptotic gain suggested by Fig.~\ref{fig:rates}.



\section{Conclusions}
\label{sec:conclusions}

We proposed a linear layered probabilistic shaping (LLPS) architecture that extends \ac{PAS} by probabilistic parity shaping (PPS). LLPS integrates with any linear \ac{FEC} and enables shaped parity bits, which are required, e.g., for optimized \ac{OOK}. \ac{LLPS} is a promising architecture for the probabilistic shaping problems considered in \cite{eriksson2017probabilistically},~\cite{git_protograph-based_2019},~\cite[Remark~3]{bocherer2019probabilistic}. The enabling component of \ac{LLPS} is a syndrome \ac{DM} (SDM) defined on a $m\times (m+\ell)$ check matrix $\mathh_p$, which maps a syndrome to the vector in the corresponding coset that minimizes a cost function. \ac{LLPS} was applied to a dirty paper coding problem, improving the energy efficiency by \SI{0.8}{dB}. Future research should develop \ac{SDM} algorithms that work efficiently also when neither $m$ nor $\ell$ are small.




\end{document}